\documentclass[twocolumn]{aastex631}
\usepackage{wrapfig}
\usepackage{amsmath}
\usepackage{graphicx} 
\usepackage{booktabs} 
\usepackage{multirow} 
\usepackage{threeparttable}
\usepackage{gensymb}

\begin{document}

\title{The Last Arecibo Message}

\author[0009-0008-3971-0022]{Kelby D. Palencia-Torres}

\affiliation{Department of Physics, University of Puerto Rico at Mayagüez, P.O Box 9000, Mayagüez, PR 00681-9000, USA}

\affiliation{Planetary Habitability Laboratory, University of Puerto Rico at Arecibo, P.O. Box 4010, Arecibo, PR 00614, USA}

\author[0000-0001-6423-6133]{César F. Quiñones-Martínez}

\affiliation{Department of Physics, University of Puerto Rico at Mayagüez, P.O Box 9000, Mayagüez, PR 00681-9000, USA}

\affiliation{Planetary Habitability Laboratory, University of Puerto Rico at Arecibo, P.O. Box 4010, Arecibo, PR 00614, USA}

\author[0009-0002-9430-594X]{Javier A. García Sepúlveda}

\affiliation{Department of Physics, University of Puerto Rico at Mayagüez, P.O Box 9000, Mayagüez, PR 00681-9000, USA}

\affiliation{Planetary Habitability Laboratory, University of Puerto Rico at Arecibo, P.O. Box 4010, Arecibo, PR 00614, USA}

\author[0009-0000-1561-8362]{Luis R. Rivera Gabriel}

\affiliation{Department of Geology, University of Puerto Rico at Mayagüez, P.O Box 9000, Mayagüez, PR 00681-9000, USA}

\affiliation{Planetary Habitability Laboratory, University of Puerto Rico at Arecibo, P.O. Box 4010, Arecibo, PR 00614, USA}

\author[0009-0008-4076-3092]{Lizmarie Mateo Roubert}

\affiliation{Department of Physics, University of Puerto Rico at Mayagüez, P.O Box 9000, Mayagüez, PR 00681-9000, USA}

\affiliation{Planetary Habitability Laboratory, University of Puerto Rico at Arecibo, P.O. Box 4010, Arecibo, PR 00614, USA}

\author[0000-0002-0601-6754]{Germán Vázquez Pérez}

\affiliation{Department of Physics, University of Puerto Rico at Mayagüez, P.O Box 9000, Mayagüez, PR 00681-9000, USA}

\affiliation{Planetary Habitability Laboratory, University of Puerto Rico at Arecibo, P.O. Box 4010, Arecibo, PR 00614, USA}

\author[0000-0003-0726-0748]{Abel Méndez}

\affiliation{Planetary Habitability Laboratory, University of Puerto Rico at Arecibo, P.O. Box 4010, Arecibo, PR 00614, USA}

\begin{abstract}

The Arecibo Message was a brief binary-encoded communication transmitted into space from the Arecibo Observatory on November 16, 1974, intended to demonstrate human technological prowess. In late 2018, to commemorate the 45\textsuperscript{th} anniversary of this message, the Arecibo Observatory initiated the New Arecibo Message competition. Following a series of challenges, our Boriken Voyagers team was recognized as the winner of the competition in August 2020. Although the primary objective of the competition was to conceptualize rather than transmit a message, the collapse of the Arecibo Telescope in December 2020 precluded any subsequent transmission efforts. Therefore, to commemorate the 50\textsuperscript{th} anniversary of the Arecibo Message, this paper presents the \textit{Last Arecibo Message}, as originally developed for the Arecibo Telescope. If the original message says \textit{we are a form of life reaching out to connect}, our message says \textit{we are ready to explore the universe together}. The prospect of transmitting this or a similar message remains an open question.

\end{abstract}

\keywords{Arecibo Message --- Interstellar Communication --- SETI --- METI --- Technosignatures --- Radio Astronomy --- History of astronomy}

\section{Introduction} \label{sec:intro}

On November 16\textsuperscript{th}, 1974, Frank Drake and the staff at the Arecibo Observatory sent the most powerful broadcast ever pointed into deep space at that time \citep{1975Icar...26..462.}. This secured their legacy not solely in historical records and scholarly articles, but also in the consciousness of everyone seeking solutions to humanity's most pressing inquiries. Since the original message was broadcast, technology has progressed in multiple ways.

Today, the world is more connected than ever. We constantly use radio frequencies to communicate with others worldwide, making them a cornerstone of our modern society. For other civilizations in the galaxy, radio transmissions might be just as important. However, we have not detected any sign of extraterrestrial intelligence. Maybe, in order to find extraterrestrials, we have to establish contact first. 

Any message meant to establish contact with extraterrestrial civilizations must undergo some scrutiny to overcome the challenges of interstellar communication. The content, the possibility of being detected, and the risk associated with sending a message must be addressed before being transmitted. This seemingly simple concept makes this concept challenging, as the receiving side must be able to decode the information we send them. Some assumptions must also be taken to model the message in a way that is universal and easier to translate and detect.

Our Boriken\footnote{Boriken is the indigenous given name to Puerto Rico by the aboriginal Taino/Arawakan people.} Voyagers team proposed a message capable of being decoded based on the constraints and assumptions required for the \textit{New Arecibo Message} challenge in 2018. In this paper, we describe our thought process for the content of the message, as well as a target visible from the Arecibo Telescope, another of the requirements of the competition. Since the telescope collapsed in December 2020, this was the last message developed by the Arecibo Observatory.

\section{Destination} \label{sec:Destination}

The purpose of our Last Arecibo Message was to send general information about Earth and the Solar System to a specific star system in the galaxy in hopes of reaching out to other intelligent civilizations. For our target, we searched for the nearest exoplanets with the best probability of sustaining habitable worlds within the transmitting declination of the Arecibo Telescope. The telescope suspended platform allowed observations within an approximate 40$\degree$ cone around the local zenith \citep{2002ASPC..278....1A}, covering a declination range from -1$\degree$ to +37.5$\degree$ \citep{2021arXiv210301367A}.

\begin{deluxetable*}{lcccccccc}
\tablecaption{Properties of stellar systems considered to receive the Last Arecibo Message. \label{table:star_system}}
\tablehead{
\colhead{Star Name} & \colhead{Type} & \colhead{d (ly)} & \colhead{$R_{\odot}$} & \colhead{$M_{\odot}$} & \colhead{$\alpha$ (deg)} & \colhead{$\delta$ (deg)} & \colhead{Age (Gyr)} & \colhead{$T_{eff}$ (K)}
}
\startdata
Teegarden's Star & M-Dwarf & $12.495$ & $0.107$ $\pm$ $0.004$ & $0.089$ $\pm$ $0.009$ & $043.25372$ & $+16.88129$ & $>$ $8$ & $2904$ $\pm$ $51$ \\
K2-288 B & M-Dwarf & $214$ & $0.32$ $\pm$ $0.03$ & $0.33$ $\pm$ $0.02$ & $055.44344$ & $+18.26881$ & $>$ 1 & $3341$ $\pm$ $276$ \\ 
K2-136 & S-Type & $191.62$ & $0.677$ $\pm$ $0.027$ & $0.742$ $\pm$ $0.03$ & $067.41247$ & $+22.88272$ & $650$ $\pm$ $70$ & $4500$ $\pm$ $125$ \\
\enddata
\tablecomments{\cite{HabitableWorldsCatalog_2024, NASA-ExoplanetArchive, 2023AJ....165..235M}}.
\end{deluxetable*}

Within this cone, the region between a declination of $15$°-$25$° has the maximum tracking time of up to $2$ hours $46$ minutes\citep{AnAstronomer'sGuideToTheArecibo305mTelescope}, giving us a large window of opportunity to transmit our message. The sequence of preparation to transmit the message adjusts our transmission window to approximately $2$ hours $30$ minutes. Therefore, our team established that, in order to complete our main objective of finding the ideal destination for our Last Arecibo Message, the Arecibo Observatory would aim to send the message to a specific stellar system within this cone. 

Our team narrowed down five different planetary systems (shown in table \ref{table:star_system} and \ref{table:planet_system}) located within the established declination range with respect to the Arecibo Observatory. Among the different candidates, our team concluded that the best destination to receive our message would be a solar system known as Teegarden’s Star. This star is located $12.5$ light years away from Earth \citep{2024A&A...684A.117D}, and its coordinates are +$16$° $51$’ $53.65$” ($16.864903$°) in declination, and at $02:53:04.59$ in the right ascension \citep{2019A&A...627A..49Z}. Teegarden's Star is an M-type dwarf star about $.107$ $R_{\odot}$ in radius and 0.089 $M_{\odot}$ in mass, the star has two planets orbiting its habitable zone, \textit{Teegarden b} and \textit{Teegarden c}, which orbit at $0.025$ $AU$ and $0.044$ $AU$ respectively from their star \citep{HabitableWorldsCatalog_2024}. These planets are of interest as they are of similar mass to Earth, however, as their star is a red dwarf, they present unique challenges to any form of life they may harbor.

Treating \textit{Teegarden b} and \textit{Teegarden c} as perfect black bodies without atmospheres and albedo of $0$, one can roughly estimate their surface temperature as $288$ $K$ and $214$ $K$,respectively, while Earth's temperature would be $278$ $K$ when approximated to a perfect black body. Since the albedo and atmosphere for the Teegarden planets have not been determined, this estimate value helps us to create a rough comparison with Earth's temperature.

Based on the information obtained from the Habitable World Catalog created by the Planetary Habitability Laboratory, one can determine each object's Earth Similarity Index (ESI), a value that can be used to compare multiple planetary properties and can lead to systematic exoplanet comparisons \citep{2011AsBio..11.1041S}. The Earth Similarity Index can be calculated for most exoplanets, from their radios or mass, and stellar flux. \textit{Teegarden c} has an ESI of $0.66$, while \textit{Teegarden b} has an ESI of $0.97$ \citep{HabitableWorldsCatalog_2024}. Therefore, \textit{Teegarden b} can be considered to be the most similar to Earth in mass and stellar flux, which also makes it an interesting candidate for the search for life \citep{Astrobiology}. These characteristics makes the Teegarden system a promising candidate to send interstellar communications and as a testing target for future messages that could reach deeper into interstellar space. Additionally, choosing Teegarden's Star benefits the quality of our message. As the star is relatively close to Earth compared to the other systems, the message will suffer little degradation as it passes through interstellar dust and gas. The transmission signal will also be very strong once it reaches the system as the message would not have lost too much power once it arrived.

\begin{deluxetable*}{llccccc}
\tablecaption{Properties of exoplanets considered to receive the Last Arecibo Message. \label{table:planet_system}}
\tablehead{
\colhead{Planet} & \colhead{Star} & \colhead{Distance to Star (AU)} & \colhead{$R_{\oplus}$} & \colhead{$M_{\oplus}$} & \colhead{$T_{eq}$ (K)} & \colhead{ESI}
}
\startdata
Teegarden b & Teegarden's Star & 0.0259 & 1.00 & 1.05 & 288 & 0.97 \\
Teegarden c & Teegarden's Star & 0.0455 & 1.00 & 1.11 & 214 & 0.66 \\
K2-288B b & K2-288B & 0.164 & 1.9 & 4.27 & 233 & 0.65 \\
K2-136 b & K2-136 & 0.0707 & 1.01 & 4.30 & 560 & - \\
K2-136 c & K2-136 & 0.1185 & 3.00 & 18.1 & 440 & - \\
K2-136 d & K2-136 & 0.1538 & 1.57 & 3.00 & 380 & - \\
\enddata
\tablecomments{\cite{HabitableWorldsCatalog_2024, 2023AJ....165..235M, NASA-ExoplanetCatalog, NASA-ExoplanetArchive}.}
\end{deluxetable*}

Another candidate for sending our message was the K2-288B system, located 214 light years from Earth. K2-288B b is an exoplanet also found in the habitable zone of its home star with a mass of 4.3 $M_{\odot}$ and a radius of 1.9 $R_{\odot}$. This Super Earth or Mini Neptune exoplanet has a black body temperature of 233 K, with an ESI of 0.65 \citep{HabitableWorldsCatalog_2024} far lower than \textit{Teegarden b}’s and not the best recipient of our message. 

The final system we investigated was the K2-136 system, 192 light years from Earth. For the three exoplanets in the system, assuming an albedo value of 0.5, K2-136 b, K2-136 c and K2-136 d have equilibrium temperatures of around 560 K, 440 K, and 380 K, respectively \citep{2023AJ....165..235M}. These temperatures and the distance of these two star systems from our planet made them poor candidates for sending our message.
   
Similarly to the original Arecibo Message, we also considered open clusters as targets to send the message because it could reach multiple stars with a single transmission. Table \ref{tab:cluster} shows the options we considered. The first option was Melotte 25, also known as Hyades. It is the closest open cluster to the Sun at approximately 150 light years away and contains about 700 stars \citep{2019A&A...623A..35L} including K2-136. Within this cluster, there are no other known exoplanets other than around K2-136, which, as previously mentioned, are not comparable to \textit{Teegarden b}.

NGC 2632, The Beehive Cluster 600 light years away \citep{NASA-HubbleMessierCatalog}, was also considered for its higher star density, but was quickly dismissed as its population consists of red giants and white dwarfs. Furthermore, by the time our message reached these destinations, it could have deteriorated in signal quality, with a low probability of being decoded.

\begin{deluxetable*}{lccccc}
\tablecaption{Properties of open star clusters considered to receive the Last Arecibo Message. \label{tab:cluster}}
\tablehead{
\colhead{Cluster Name} & \colhead{$d$ (ly)} & \colhead{$\alpha$ (deg)} & \colhead{$\delta$ (deg)} & \colhead{Radius (ly)} & \colhead{Star Population}
}
\startdata
Melotte 25 (The Hyades) & 150 & 067.4470 & +16.9480 & 10 & $>$700 \\
NGC 2632 (Beehive Cluster) & 600 & 130.0540 & +19.6210 & 11 & $>$1000 \\
\enddata
\end{deluxetable*}

\begin{figure*}[ht!]
\plotone{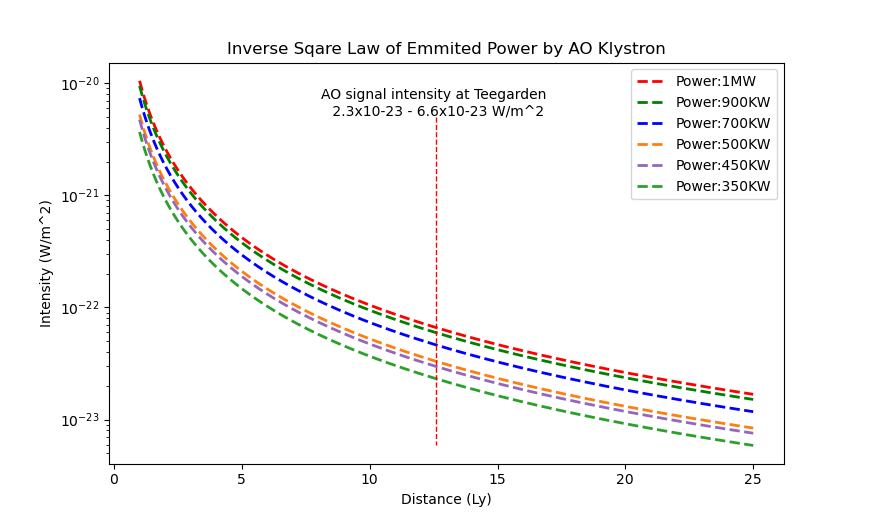}
\caption{Intensity reduction as a function of distance of the power output capabilities of the Arecibo Observatory's klystrons.
\label{fig:PowerAO}}
\end{figure*}
    
Considering the distance of our candidates and the possibility for recipients to detect the message, Teegarden b was selected as our target. Being the closest planet from our list, its recipients might be able to detect the signal emitted by the Arecibo Telescope. In addition, its high ESI means that it is similar to Earth's mass and insolation.
    
In theory, the S-band transmitter of the Arecibo Telescope in the 1970s was able to generate an output power of 1 MW, with an EIRP of 25 TW \cite{Arecibo_Operation_and_Maintenance_Manual}, but its maximum power output has been recorded to be 900 kW. In 2020 the Arecibo Observatory had one functioning Klystron that has been measured to emit from 350 kW to 450 kW of power output. This meant that the farther we wanted to send the message, the more powerful the receivers on the destination must be.

We plotted what the power would be if it was received at 12.495 light years from Earth, assuming that the receivers would have technology similar to that of the Arecibo Observatory. As can be seen in figure \ref{fig:PowerAO}, the intensity decreases by the inverse square law. This shows that the power they will receive per square area is quite small. For our particular candidate, if we consider the power output of the conditions in 2020 (350 kW and 450 kW), we obtain an intensity of approximately $2.98\times10^{-23}$ Wm$^{2}$ to $2.32\times10^{-23}$ Wm$^{2}$.

\begin{figure*}[ht!]
\plotone{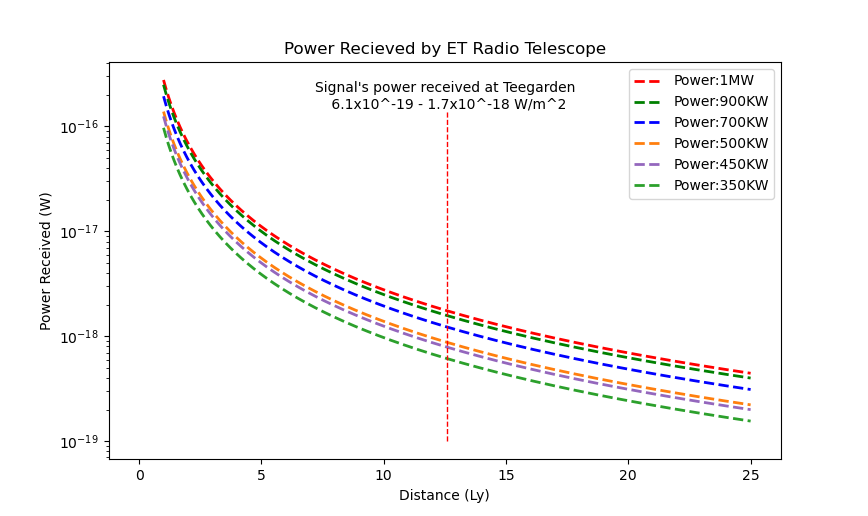}
\caption{Power received from a radio telescope similar to the Arecibo Telescope at the distance of Teegarden's Star. 
\label{fig:PowerET}}
\end{figure*}
    
If we considered the receivers to have the same effective area of the observatory of 26,414.8 m$^2$ then they would receive a power of at least 7.87$\times10^{-19}$ W to 6.131$\times10^{-19}$ W as seen in figure \ref{fig:PowerET}. Compared to its full functioning capacity of 1 MW or 900 kW the signal received would have one less order of magnitude at 1$\times10^{-18}$ W. This is extremely weak compared to the sensitivity of the Arecibo Observatory’s capabilities if compared to detectable Pulsars in S band with a flux of 0.1 Jy and bandwidth of 100 MHz. That means that the technology needed would have to be more sensitive than ours in order to detect our message, unless they have a bigger dish or array of radio telescopes.

\subsection{Recent Teegarden's Star Developments}

New studies have revised the parameters for Teegarden exoplanets. For example, in early 2024, the study "Teegarden's Star revisited: A nearby planetary system with at least three exoplanets". The team led by Stefan Dreizler revised the orbital parameters of Teegarden b and Teegarden c with new radial velocity measurements from CARMENES, ESPRESSO, MAROON-X, and HPF with photometry measurements from TESS \citep{2024A&A...684A.117D}.

These refinements have changed the ESI values for the exoplanets with Teegarden b from 0.97 to 0.90 and Teegarden c with a larger increase from 0.66 to 0.88. The study also discovered a third exoplanet, Teegarden d with a mass of 0.82 M$_{\oplus}$ and an orbital period of 26.13 days with am equilibrium temperature of 159 K. Unlike the other two exoplanets in this stellar system, Teegarden d orbits outside of the habitable zone and is not a viable candidate for our project. Teegarden b no longer holds the highest ESI ranking in the Habitable Worlds Catalog. It is important to keep in mind that the ESI is not a measure of habitability; therefore, our team still considers Teegarden's Star as the best candidate to receive our message. 

\subsection{Where to point?}

To send the message, we need to consider the proper motion of the selected candidate. Due to the distance, the time it takes light to arrive at Earth is 12.495 years, meaning that what we are observing today is 12.495 years in the past. Today, that system must be in another position 12.495 years ahead of where it is currently seen. Since our message will travel at that time the position would be 24.99 years ahead for it to receive the message. The system is at 16° 52' 52.64'' in declination and 2h 53m 0.89s in right ascension and has a proper motion of 3429.0828 masyr and  -3805.5411 masyr \cite{2021A&A...649A...1G}. For those 24.99 years, the system would have moved to 16° 50' 1.42" and 2h 53m 11.34s. This is the location the message should be sent for it to arrive at the Teegarden system in 12.495 years from now.

\section{Technology} \label{sec:Technology}

\subsection{Signal Specifications}

Before our message is sent into space, we had to consider the broad range of frequencies that we can use and the uncertainty of what other intelligent civilizations will be listening to. A good assumption is that other civilizations will be monitoring and listening in the 1420-1662 MHz "water hole" range \citep{morrison1979search}. It would be ideal to use the 21 cm wavelength, from the hydrogen hyperfine transition, to send the message. However, this is a protected frequency allocated for Radio Astronomy that can not be used for transmissions. 

Although radio telescopes can detect signals in this range, Arecibo did not possess a transmitter at this frequency. The Arecibo Observatory had three available transmitters. We proposed to use the capabilities of Arecibo's S-band Planetary radar transmitter in the 2370-2390 MHz range. This S-band frequency range is ideal for our objective due to its ability to pass through Earth-like atmospheres and cosmic dust across interstellar distances with little signal degradation.
        
We proposed using the same frequency modulations of 10 Hz as the original message to distinguish between ones and zeros. The small frequency modulations mitigate the risks that our signal falls outside the objective frequency range of observation. This frequency modulation is also implemented in the prephase and endphase of the message, in which we seek the attention of our objective and mark the end of the message. The prephase is divided into three sets of bits, which could be emitted at 20 bits/s. The prephase and the endphase last approximately 28 seconds each and is explained in Section \ref{sec:Message Content}. The second segment of the message contains 5,402 bits, which at 20 bits/s, takes approximately 4.5 minutes to send. In summary, The Last Arecibo Message could be transmitted in approximately 5.5 minutes. The transmission can be repeated up to a maximum of 27 times, which will take a total of 148.5 minutes. These repetitions would maximize the 2.5 hour observation window that the Arecibo Observatory had.
        
\subsection{Energy Consumption}

One essential aspect that we need to consider to send our message is the energy required to power up the whole process. Some rough estimates were calculated regarding the power usage of the observatory. We then divided the energy consumption into three stages: Stage 1 - Powering and warming up the transmitter, Stage 2 - Sending the signal, and Stage 3 - Moving the telescope.
        
Stage 3 runs on commercial energy but stages 1 and 2 consume energy produced by diesel generators, which is then redirected to the available klystrons. We focus on the energy consumption and the diesel requirements during stage 2, sending the message. The maximum efficiency of the radar klystron is 500 kW. However, due to the 50$\%$  efficiency of the system, we are required to provide fuel to generate approximately 1 MW if we want to send our signal at peak efficiency.
        
Stage 2 is divided in three parts, the prephase, the content, and the endphase. The prephase and end phase will last a total of 0.92 minutes and the content takes 4.59 minutes. The Planetary Radar consumes approximately 2 gallons of diesel per minute on average. Therefore, we would require approximately 11.02 gallons of diesel for the 5.51 minutes that the message would last in each repetition; in contrast to approximately 22.07 gallons that would require if we used a rate of 10 bits/s to send the signal. Our team chose to use a rate of 20 bits/s to decrease this required energy and mitigate the consequences on our environment due to the consumption of fossil fuels.
        
If we choose to send the message continuously for 148.9 minutes during the 2.5 hour window, then it would consume approximately 297.8 gallons of diesel. Considering that this may greatly affect the environment, less repetitions are recommended, but it would negatively impact the chances of our signal being detected.

\section{Message} \label{sec:Message}
 
\subsection{Assumptions}

Before we created our message, we considered the conditions that must be met for successful contact with another intelligent civilization. Currently, there are too many unknowns to determine if our message will succeed since we have yet to received a signal from intelligent lifeforms. That is, for another intelligent civilization, it would be equally challenging to receive our transmission. This civilization might understand concepts such as mathematics and language differently, thus also affecting the way their civilization developed. In the case of successful contact with a civilization, where there could be a chance of receiving a response, such a civilization would have to fulfill the following assumptions.
    
\subsubsection{Knowledge of Various Scientific Concepts}

The civilization must have achieved a deep understanding of various fields of the various fields within science, electromagnetism, optics, engineering, and mathematics to develop the tools and machinery necessary to detect and understand the message. Any civilization that has not developed these concepts will not be prepared to detect our message as it reaches their stellar system.
            
\subsubsection{Harnessing Electricity}

Having this knowledge, the civilization would be able to progress its technology through periods similar to those that humanity underwent in the past centuries. Technological progress would eventually lead this civilization to harness electricity allowing them to create computers and wireless communication systems. For this to happen somewhere else in the galaxy, these civilizations must accomplish the first assumption. Without achieving this level of development, our message will not be received and decoded.
            
\subsubsection{Development of Multi-wavelength Astronomy}

For as long as humans have looked at the stars, we have aspired to know what lies beyond our planet. At first, we could only study what we could see with our eyes, limited to as far as our largest optical equipment could allow us to see. Today, we are able to observe the cosmos using most of the electromagnetic spectrum and so must any other who wants to communicate with another sentient species. For the purpose of our message, any civilization that could encounter it would need to be observing on the S-band frequency range and be pointed at the general direction of Earth. Thus, any civilization must have developed multiwavelength astronomy to detect it, particularly in the radio section. This means that the receivers must have a reason as to why they study the radio frequency, specially the S-band frequency that will be used for the message. For this to happen, the second assumption must be met, or the message will not be received or decoded.
        
\subsubsection{Analogous Technological Development}

Any civilization that meets the mentioned assumptions and meets the required technological advancement will need to be on the path of the radio telescope. Finally, once the civilization receives our message and decodes it, they might want to send a response. At minimum, we expect them to have reach a technological level on par with the capabilities of the Arecibo Telescope. If the recipients lie farther away, they would need much more powerful radio telescopes to ensure the signal integrity. If they do not posses the necessary equipment to detect faint signals or have a better telescope than that of the Arecibo Observatory they might receive the signal but not be able to recognize it.
            
\subsection{Message Content} \label{sec:Message Content}
        
\subsubsection{Prephase and Endphase}

\begin{figure}[ht!]
            \centering
                \includegraphics[width=1\linewidth]{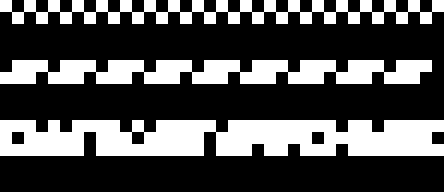}
                 \caption{Prephase and endphase of the message used to call for attention and to indicate that the message ended.
                 \label{fig:AO_prephase}}
            \end{figure}

The message is divided into three parts: the prephase, the main message, and the endphase. The prephase is divided into three sets of bits with different patterns shifting 10 Hz from the main frequency of 2380 MHz (Figure \ref{fig:AO_prephase}). Each set will be separated by a 5.55-second interval. The first set will be a consecutive shift from the main frequency that repeats 37 times with the same time intervals (i.e., 20 bits/s). This pattern will be used to call the attention of the listeners and will include the length of the message. The length is a prime number to help decipher the image created by the message.

The second set is a repeating pattern that shifts 10 Hz from the main frequency for 3 bits and then shifts back to the main frequency, i.e., '011101110'. This signal indicates that it is without a doubt artificial. The last set will include the number pi, with each group of numbers representing the digits, i.e 3.14... = '011101011110...'. The prephase and the endphase have the same content.

\subsubsection{Visual Representation of the Message}

            \begin{figure}[ht!]
            \centering
                \includegraphics[width=0.62\linewidth,height=21cm]{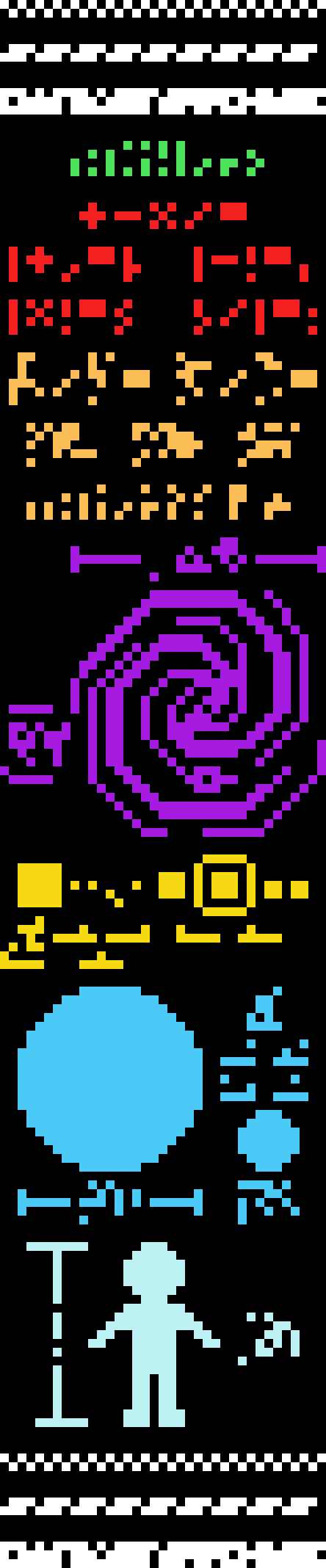}
                 \caption{The Last Arecibo Message is divided into seven sections (colored areas) with different components to decode. 
                 \label{fig:AO_message}}
            \end{figure}

\subsection{Meaning of the Message}

The main message is a 37 $\times$ 149 bit grid that includes some of the information sent from the original message (Figure \ref{fig:AO_message}). We think that this information is the key to our message and must be included, since it encompasses fundamental knowledge about our species, planet, and sciences. Some of the original components of the original Arecibo Message were updated in this new message. For example, the message will feature additional mathematical and physical information, as well as numbers encoded in binary that are not written horizontally, if not vertically, to keep consistency. The main message is divided into seven sections as follows:

\subsubsection{Numbers from 1 to 10}

This section represents the numbers from 1 to 10 in binary form. Our team decided to include them in the Last Arecibo Message, in the same case as the previous Arecibo Message \citep{1975Icar...26..462.}, with the objective of showing the base of our numerical system, as well as denoting how to decode the message, as it contains the least significant digit marker. If the recipient of this message were able to decode this section of the message and understand its significance, we can imply that they would also be able to decode the remainder of the message since they would also have the ability to comprehend the system we have designed to represent this information.

\subsubsection{Arithmetic Symbols and its functions}

This section holds a visual representation of our basic arithmetic operators of addition, subtraction, multiplication, division, and equality. We decided to include four basic arithmetical operations as part of the message's content, with the main objective of demonstrating the function of each operator. The included operations are the following:
\begin{align*}
7+8=15 && 7-6=1 && 7 \times 6=42 && 14/7 = 2.
\end{align*}
By adding these operations and showing the function and usage of each of the operators, it will help us include even more numbers in its content, and it will also allow us to share some of our mathematical knowledge on how we understand the universe.

\subsubsection{Mathematical and Physical Constants}

This section holds one of the most valued constants that helps explain how we understand and see the world, which is of great significance in the world of science. In a similar way as the arithmetic symbols and their function section of the message, we constructed mathematical problems for the recipient of the Last Arecibo Message to solve. These other arithmetic functions include $\pi$, approximated by dividing 2 integers (in this case, 52163/16604) to obtain its value, 3.141592 up to 6 decimal places, and \textit{e}, represented with 5 decimal places of accuracy by solving the operation 271828/100000 = 2.71828.

Additionally, among other important constants in the scientific community, we also included the value of the speed of light as 299,792,458 m/s, and the Planck and Boltzmann constants. In the case of these two latter constants, they were written as integers with their respective eight decimal places. Lastly, our Last Arecibo Message also includes the famous Fibonacci Sequence until the eleventh place in the sequence.
            
\subsubsection{Milky Way Galaxy and our Location}

In this section, we include a visual aid of the Milky Way galaxy to scale, in hopes of sparking the message recipient's curiosity. If the assumption is indeed correct and they manage to recognize the figure as another galaxy, they as well must be observing other galaxies and understand that they are located within one. Additionally, the recipients should be able to understand that galaxies are a common astronomical object in the universe and they should have a method for measuring their distances.

In the cases of the figure included in the message, each bit in the image represents approximately 1 kpc in distance. There are two scale bars in the image, its information being represented in light years one along the horizontal axis that indicates the diameter of the galaxy, and a vertical one that indicates the time it takes the light to travel from the galactic center to the Sun (in this case, it being around 8.15 kpc from its center) \citep{2019ApJ...885..131R}. Additionally, this last bar also indicates, with a square around a 0-bit, the location of our Solar System. 
            
The reason we chose light year units to represent this information is because, when multiplied by the seconds in a year, in example, the result can also indicate how long it takes for the light to travel through a specific distance. By presenting it in that way, the recipients of the message can have the ability to multiply the number by the speed of light (i.e., frequency times the wavelength of the signal) to obtain the indicated distance in meters. In the case of the distance from the galactic center to the Solar System, the number in binary is 841,159,728,000 and for the diameter of the Milky Way, the number is 3,153,600,000,000. Both, when multiplied by 2380 MHz and 0.1259632176 m, would result in distances of 2.52$\times10^{20}$ m and 9.45$\times10^{20}$ m respectively.
            
\subsubsection{Solar System and the 8 planets}

This section is a visual representation of the solar system. This is an updated version from the original message, where we included the Moon next to Earth, added rings to Saturn, and adjusted the size of the planets for scale. The Earth is shifted down from the line center of the system to indicate that this is the planet from which the message originated. Below the solar system we included a similar design of the elements like in the first message, but to indicate the most abundant elements in the planets, dividing them into rocky planets and gaseous planets. Three elements were chosen to indicate the main composition of the rocky planets: oxygen, silicon and iron \citep{Neser2022}. Similarly, Hydrogen and Helium represent the gaseous planets. All elements in this section are grouped together under their corresponding planets. 
            
\subsubsection{Earth-Moon System}

This section of the message is another visual aid of the Earth-Moon system indicating the diameter of each body and some compounds in the atmosphere of the Earth. The elements included are hydrogen, carbon, nitrogen, and oxygen. With the identified elements, the following molecules are represented by O$_2$, CO$_2$, H$_2$O and N$_2$ in the same order. The diameter of the Earth is represented with the number in binary 101,156,514 and if multiplied by the wavelength, you get the diameter of the Earth in meters (12,742,000 m). Similarly for the Moon, its diameter is represented by the number 27,581,067 and, if multiplied by the wavelength, we can obtain its diameter in meters (3,474,200 m).
            
\subsubsection{Human Representation}

This section is the last section of the message and includes a visual aid of a humanoid figure with more detail, compared to the original message. The average height of humans (1.7 meters) is represented by the number 14 in binary, and if multiplied by the wavelength, you get the height in meters. On its side, there are binary digits that represent the human population of 7.8 billion humans in 2020 (today 8 billion). After this message finishes, the endphase (i.e., the same as the prephase) starts informing the receiver that the message has ended.
  
\subsection{Risks}

Communicating with extraterrestrial life is a highly debated subject with many interpretations. The results of making contact could be detrimental to humanity and should be taken with extreme caution. Within the debate, there are two main sides, those who believe that humanity should remain as listeners to extraterrestrial communications and those who want humanity to actively reach out and establish contact.

The first group is embodied by the Search for Extraterrestrial Intelligence (SETI). Their concern about sending a message from Earth has the possibility of attracting unwanted attention by hostile extraterrestrials who might act against our planet in ways that might be unknown to humans \citep{SETI_STATEMENT}. If our message is verbose in sensitive details about Earth, it could leave humanity vulnerable to a technologically superior civilization.

The other group is represented by Messaging Extraterrestrial Intelligence (METI), who argue that beaming information into the cosmos is a necessary practice in the efforts to discover extraterrestrial intelligence. Proponents of METI dispute, among other things, \cite{BeamedSignal} that humans have been sending messages to outer space in the form of transmissions from TV stations, radio and other forms of wireless communication. Although some of the early transmissions are now venturing far from Earth, these transmissions might not be detectable by another civilization due to signal degradation and signal power loss. Nevertheless, the fact is that we have been exposing ourselves for decades now and we will most likely continue to do so, as technology progresses.
            
Similarly, questions have been raised as to who has the right to send these messages and what content should be included in them. The Last Arecibo Message takes inspiration from both SETI and METI. We tried to make the message not too revealing about Earth as a precaution, according to SETI Statement values. When proposing what goes into the message itself, we identified potential risks that would need to be mitigated to be considered in the transmission.
     
\subsubsection{Technical Risks}

\textbf{Signal is not delivered to the intended destination:}
This risk has been addressed by carefully choosing the destination of the message using appropriate data and parameters such as declination, right ascension, and distance from our solar system, as previously mentioned in the objective location. If the signal is not delivered to the chosen target, there will probably be no repercussions. 
            
\textbf{Signal is intercepted and a response is received:}
To address this risk, we have concluded that the optimal mitigation for this is to inform the international community about the message, when it is sent, where it is being sent, and the content and purpose of the message. Informing other nations about this new message will prepare us in the event that the message is intercepted and a response is received. 
            
\textbf{Excessive disclosure of information about our species and our planet:}
This is an evident risk that cannot be entirely controlled if the objective is to contact intelligent civilizations and show them information about ourselves and our place in the universe. This was addressed to some extent by including general information that can be gathered by observing and understanding the universe. These mathematical concepts and physical constants should already be known to any civilization capable of detecting the message.
            
\textbf{Decoding the message:}
The message is encoded using binary notation and it is assumed that an intelligent civilization can decode it for its mathematical simplicity. This binary format also allows the message to be visually interpreted through pixel imagery. Assuming that extraterrestrials have similar visual capabilities and pattern recognition skills, this visual representation can resolve discrepancies in their interpretation of the message. However, it is assumed that if a civilization is capable of receiving this message, they would know the universal constants encoded in the message that must be used to decode the message.
            
\textbf{They might never receive the message because they would need to observe at a specific time:}
The message includes a 'beacon' to draw the attention of receivers of this signal. This will serve as the announcement for the full message that will arrive later. But regardless of the said 'beacon', the receiver should be surveying the sky at a specific time in our direction in their night sky. Subsequently, their planet's orbital period and rotational period can significantly impact their ability to receive our message. This could be true for Teegarden b since its rotational period is not known.
            
\subsubsection{Other risks}

\textbf{Misinterpretation of the message:}
Our team did not want to produce a message that presented our species as hostile or as a colonizer of worlds. Therefore, after careful consideration, the team chose not to include any signs of our capability for spaceflight. A pictogram that depicts this capability could be misunderstood and taken as a threat to invade other worlds. 
            
\textbf{Assumptions of intelligent civilization technology:}
As the intensity of a signal decays as it propagates through space per unit of area traveled, the farther the destination, the harder it is to detect our message. This means that intelligent civilization would need sophisticated technology that would be sensitive enough to detect our signal. This implies that listeners would probably need more advanced technology than what we currently have on Earth. We made the assumption that intelligent civilization should have a radio telescope with the shape and diameter 21.36 times larger than the 305 m diameter of the Arecibo Telescope. Furthermore, this would also imply that the intelligent civilization has significant engineering knowledge and capabilities to build a structure of 133,337,420 m$^2$ in total size.
            
\textbf{Biological limitation of the recipients of the message:} 
We cannot assume the extraterrestrials will share any of our basic senses, this could pose a limitation when interpreting the message itself. We also face the possibility that the recipient species may not have a sense of hearing or vision in the same way as us, posing a challenge to interpreting the message as it is completely dependent on visual imagery.
            
\textbf{Habitability of Teegarden’s Star:}
Even though Teegarden b has a high ESI value among other studied exoplanets, it is also prone to certain types of risk that could affect its habitability. Teegarden's Star is a M-dwarf star, the most common type of star in the Milky Way Galaxy, that has Earth-sized planets in the habitable zone \citep{2020MNRAS.494.1297M}. However, M-dwarf stars are also known to generate strong flares, winds, and coronal mass ejections that can affect the exoplanets orbiting their habitable zones and their habitability \citep{2007AsBio...7..185L, 2019AsBio..19...64T, 2017ApJ...841..124V}.

Research done by \cite{2017ApJ...836L...3A} has indicated that the generation of these stellar flares and storms also leads to the generation of X-ray and extreme ultraviolet emissions at their habitable zones, and by creating a model consisting of planets orbiting red dwarf stars, it can lead to oxygen escapes on said planets \citep{NASA}. The intensity of these oxygen losses scales off the amount of energy the star emits, since the more X-ray and extreme ultraviolet energy there is, the stronger the ion escape effect becomes.

In models, hydrogen, oxygen, and nitrogen, the latter two being essential molecules of life, are able to escape the atmosphere of these planets \cite{2017ApJ...836L...3A}. Therefore, this type of scenario presented in exoplanets orbiting a red dwarf star with strong stellar winds could eliminate the planet's water supply before life was allowed to be developed \citep{NASA}.

Although there has been no confirmation of the presence of stellar winds in the atmosphere of Teegarden b, we cannot discard the possibility of Teegarden's Star generating strong winds that could potentially affect their exoplanets Teegarden b, c and d, and the possibility of them being able to develop life. Additionally, the presence of strong stellar winds could also prevent the planets in the Teegarden's Star system from developing a stable atmosphere and therefore affect their potential habitability \citep{2013A&A...557A..67V}.

\section{Conclusion} \label{sec:Conclusion}

The Last Arecibo Message was the last message designed by the Arecibo Observatory, as part of a global competition by students. It has similarities to the original Arecibo Message, but it has more mathematical and astronomical knowledge and does not include information about our biology. The message is larger and encoded in 5,513 bits. It has seven sections with numerical and visual content. It also has a prephase and an endphase with three different patterns intended to draw the attention of receivers. The former Arecibo Telescope was capable to send the message using the S-band transmitter at 2380 MHz by shifting the frequency by 10 Hz at 20 bits/s. It would have taken 4.59 minutes to transmit (without the prephase and endphase) and consume 11.02 gallons of diesel.

The message could be sent to the Teegarden's Star system in hopes that our message can reach the planets Teegarden b and Teegarden c, both located within the habitable zone. It will take approximately 12.495 years for the message to arrive at the system, and it would take around 25+ years for us to ever receive a reply, if any. Due to the distance, the potential extraterrestrial civilization would need a more sensitive radio telescope than what we currently have for detection. We are assuming that they have knowledge about harnessing electricity, multiwavelength astronomy, and analogous technological development. There are possible risks that arise by sending a message and letting the receiver know our location. An alternative will be to exclude information about our Solar System and its location in the galaxy.

The original Arecibo Message conveyed a more personal depiction of terrestrial life and human existence. In a single sentence, it can be interpreted that the intended message can be summarized as \textit{we are a form of life reaching out to connect}. In contrast, our Last Arecibo Message emphasizes universally recognized astronomical knowledge to say "we are ready to explore the universe together".

With this message, our aim is to continue and build upon the legacy established by Frank Drake with the original Arecibo Message. We hope that this message also sparks curiosity and interest in space and in the search for life in the universe. For example, it inspired artists to create a \href{https://phl.upr.edu/message/meraki}{musical score} as part of the message.

We believe that the Last Arecibo Message will shine a light on the importance of the Arecibo Observatory in the exploration and the development of space. May this message help us answer if we are truly alone in the universe, \textit{per aspera ad astra}.
    
\section{Acknowledgments}

As the Boriken Voyagers Team, we acknowledge Gaddiel González, Stephanie Silva, and Wilbert Ruperto as additional students who contributed to the team and worked in the original proposal for the competition. We were all members of SEDS-UPRM student organization and we thank them for endorsing the team. We extend our thanks to Dr. Edgard Rivera-Valentin and Dr. Robert Minchin for providing aid with information about the Arecibo Observatory and its capabilities. We thank Dr.Henri Radovan as the mentor of our project. Thanks to Dr. Alessandra Pacini for organizing the New Arecibo Message Global Challenge at the Arecibo Observatory. We acknowledge the use of the Gaia Catalog, the NASA Exoplanet Archive, and the Habitable Worlds Catalog in this project.

\bibliography{last_arecibo_message}{}
\bibliographystyle{aasjournal}
\end{document}